\documentstyle[epsfig,graphicx]{mn}

\makeatletter
\def\@cite#1#2{(\if@tempswa #2 \fi #1)}
\makeatother

\def\refeq#1{{(\ref{#1})}}
\mathcode`\|="8000		
{\catcode`\|=\active \gdef|#1{_{\rm #1}}}

\def\ee{\protect\pee}
\def\pee#1{\ifmmode{\times10^{#1}}\else$\times10^{#1}$\fi}
\def\rd{{\rm d}}                                      
\def\rdif#1#2{\mathchoice{\rd{#1}\over\rd{#2}}{\rd{#1}/\rd{#2}}
{\rd{#1}/\rd{#2}}{\rd{#1}/\rd{#2}}}                            

\def\unit#1{\,{\rm {#1}}}
\def\Kelv{\unit{K}}

\def\ion#1#2{{\rm #1}%
\ifmmode{\mathchoice{\scriptstyle}{\scriptstyle}
{\scriptscriptstyle}{\scriptscriptstyle}{\rm\uppercase{#2}}}%
\else$\,\scriptstyle\rm\uppercase{#2}$\fi}

\def\HII{{\ion{H}{II}}}
\def\etal{{et~al.}}
\def\eg{{e.g.}}
\def\ie{{i.e.}}

\def\cf{{cf.}}

\title{Resolved shocks in clumpy media}

\author[R.J.R. Williams \& J.E. Dyson]{R.J.R. Williams$^1$ and J.E. Dyson$^2$\\
$^1$Department of Physics and Astronomy, Cardiff University, PO Box 913,
Cardiff CF24 3YB\\
$^2$Department of Physics and Astronomy, University of Leeds, Leeds LS2 9JT
}
\date{Received **INSERT**; in original form **INSERT**}
\pagerange{\pageref{firstpage}--\pageref{lastpage}}
\begin{document}
\label{firstpage}
\maketitle

\begin{abstract}
We study the structure of shocks in clumpy media, using a multifluid
formalism.  As expected, shocks broaden as they weaken: for
sufficiently weak shocks, no viscous subshock appears in the
structure.  This has significant implications for the survival of
dense clouds in regions overrun by shocks in a wide range of
astrophysical circumstances, from planetary nebulae to the nuclei of
starburst galaxies.
\end{abstract}


\section{Introduction}

Astrophysical media are often strongly clumped, with dense clouds or
bullets of cool gas distributed in a far hotter external medium.  The
flows in these media are complex, particularly when they are
perturbed, for example, by outflows from nearby young stars or
supernovae.

Insight into these structures be gained by approximate treatments,
including multiphase hydrodynamics.  Many mass-, momentum- and
energy-transfer processes between phases will be important in real
flows \cite{shuea72,drew83,kamaya97b}; similar treatments are used for
multiphase space plasma flows, as recently reviewed by Szeg\"o
\etal{}~\shortcite{szego00}.  The simplest cases of pure mass-loading
or momentum transfer are interesting limits.

The mass-loading approximation treats the dense cool phase as a
distributed source of matter within a hot matrix.  It is assumed in
this treatment that subsequent to the mass-loading, dissipative
processes such as thermal conduction, enhanced by turbulent mixing,
mean that the hot phase can be treated as a single, smooth flow.  This
approximation has been applied, by ourselves and others, to diverse
astrophysical objects, such as Wolf-Rayet and planetary nebulae,
ultracompact \HII\ regions, starburst galaxies and active galactic
nuclei \cite{cc85,hdps,adh94,redea96,smith96,wilbp97}. 

Underlying these approximations are the details of the interaction
between a dense clump and a steady wind or a shock have been studied
numerically by many authors \cite[\eg{}]{wood72,nfg82,kmc}.  Where a
strong shock propagates over a dense cloud, these authors find that
the cloud is crushed in the direction parallel to the incoming shock,
and expands in the lateral direction.  Whether the net effect on the
cold phase gas can better be described as destruction of the clouds or
as a redistribution of their mass spectrum depends on details of
radiative and transport processes not included in these simulations.

If the shocks are broadened substantially, the impinging hot material
may be treated as a gradually accelerating stream.  At least in
radiative gas, steady stream interactions may be less destructive than
those with rapidly accelerating shocks \cite{schi95}, as a result of
sacrificial stripping of small instabilities and of the dissipation of
frictional heat input.  Indeed, shocks may be able to cause the
agglomeration of the condensed material into larger clumps
\cite{kamaya97b}.  The increased pressure of the diffuse gas can also
lead to condensation of material into the dense phase, although this
depends on details of the thermal evolution.

The results of an interaction with many sources has in general been
treated in a continuum approximation: simulations in which the
interaction of the flow over numerous obstacles is studied in detail
are only just becoming feasible \cite{jjn,polud01}, and the number of
obstacles in these studies is limited.  In Fig.~1 of Jun \etal{}, the
leading shock is apparently slowed and weakened somewhat by the
presence of the clouds, but it is difficult to infer the overall
behaviour of the shocks from this paper as only two simulations are
illustrated.  The finite resolution of such studies leads to numerical
viscosity and numerical thermal conduction which can result in far
more effective coupling between the phases than might be expected in
reality.

In the present paper, we treat the multiphase structure using a
momentum-loading formalism.  This allows a continuum treatment of the
internal structure to maintain the distinction between cool- and
hot-phase properties, while including both as dynamically active
components of the flow.  We parameterize the uncertainty about the
detail of the momentum-transfer process by considering separate forms
of the coupling terms, and look for general features of the results.
We study the structure of steady shocks in a multiphase ISM in this
approximation, for the case in which the two phases ahead of the shock
are comoving.  This case is somewhat distinct from the formation of
termination shocks in an impinging wind driven by mass injection from
gas which has been overrun, e.g.\ in systems such as cometary particle
injection in the solar wind~\cite{bierea67,zo91}, or clumps which have
been overrun by bubbles driven by stellar winds~\cite{whd95} or
supernovae \cite{cmo81,wd87,pittea01}.

Kamaya has studied the wave dispersion relation~\cite{kamaya97a} and
time-dependent shock tube problem \cite{kamaya97b} for two phases
initially at rest with respect to each other.  The time dependent
solutions in the latter paper were not integrated for a time
sufficient for the two phases to reach equilibrium between the shock
and the rarefaction; only one case (with a leading subshock) is
presented.  Our study complements this work, by determining
steady-state structures for a well-defined, if simplified, set of
model equations.

We discuss the implications of our results for various astrophysical
systems.

\section{Basic equations}

We consider a system of two phases, a distributed adiabatic hot phase
and a clumped cold phase which fills a negligible fraction of the
volume of the flow.  In reality there will be significant substructure
in both the hot and cold phases, but we use continuum equations for
both phases on the assumption that a suitable coarse-graining process
can provide consistently-defined local mean values for the flow
density, velocity and pressure.

We assume that the hot phase obeys mass, momentum and energy equations
\begin{eqnarray}
\rdif{}{x}\left(\rho_1 v_1\right) &=& 0 \label{e:1mass}\\
\rdif{}{x}\left(p + \rho_1 v_1^2\right) &=& {\cal S} \label{e:1mom}\\
\rdif{}{x}\left[v_1\left(\gamma p\over\gamma-1\right)+
{1\over 2}\rho_1 v_1^3\right] &=& {\cal S}\left[v_1 + \alpha(v_2-v_1)\right],
\label{e:1en}
\end{eqnarray}
where $\rho_1$ is the gas density in the hot phase, $v_1$ is the flow
velocity, and $p$ is the pressure.  The source term $\cal S$ is the
slip force between the two phases, and $\alpha$ is the fraction of the
frictional energy dissipation resulting from this slip which acts to
heat the hot phase.  In the absence of a second phase (${\cal S} =
0$), these equations are simply the equations of steady-state flow for
a fluid with adiabatic index $\gamma$.

The clumped cold phase obeys the mass and (ballistic) momentum
equations
\begin{eqnarray}
\rdif{}{x}\left(\rho_2 v_2\right) &=& 0 \label{e:2mass}\\
\rdif{}{x}\left(\rho_2 v_2^2\right) &=& -{\cal S},\label{e:2mom}
\end{eqnarray}
where $\rho_2$ is the mean density per unit volume of the cold gas,
rather than the density within individual clumps.  These equations
form an autonomous system, as $x$ appears only in the differential
operators.

A particularly simple set of equations is used here, considering only
momentum-transfer effect.  To accurately treat a specific flow, a wide
range of mass and energy-transfer processes should also be considered
\cite{shuea72,wd87}.  However, the treatment of the astrophysical
plasma as consisting of locally uniform interacting media is itself a
considerable approximation, and we will see that our further
simplification of the system leads to interesting results.

Equations~\refeq{e:1mass}, \refeq{e:2mass} and the sum of
equations~\refeq{e:1mom} and~\refeq{e:2mom} can be immediately
integrated, to give
\begin{eqnarray}
\Phi_1 &=& \rho_1 v_1\label{e:phi1}\\
\Phi_2 &=& \rho_2 v_2\label{e:phi2}\\
p+\rho_1 v_1^2 +\rho_2 v_2^2 &=& \Pi.\label{e:pi}
\end{eqnarray}
In addition, $\cal S$ can be eliminated from
equations~\refeq{e:1mass}--\refeq{e:2mom} to give
\begin{equation}
{\rdif{}{x}}\left[{\gamma p v_1\over\gamma-1}+{\alpha\over2}\left(\rho_1v_1^3
+ \rho_2v_2^3\right)\right] =
(1-\alpha)v_1\rdif{p}{x}.\label{e:pv}
\end{equation}
From this equation it is clear that for $\alpha=1$, the total energy
flux
\begin{equation}
{\cal E} = {\gamma pv_1\over\gamma-1}+{1\over 2}(\rho_1v_1^3+\rho_2v_2^3)
\end{equation}
is constant.  For $\alpha = 0$, the hot flow is adiabatic {\it so long
as there are no discontinuities in the flow}, and hence $p \propto
v_1^{-\gamma}$.

For general $\alpha$, the dependence on $x$ can easily be eliminated,
as can those on $\rho_1$, $\rho_2$ and $p$, using the
integrals~\refeq{e:phi1}--\refeq{e:pi}.  If the solution is smooth,
the result is a first-order o.d.e.
\begin{equation}
\rdif{v_1}{v_2} =
{v_1-\alpha(\gamma-1)(v_2-v_1)\over 
\gamma(\Pi/\Phi_2) - (\gamma+1)(\Phi_1/\Phi_2) v_1 - \gamma v_2}.
\label{e:v1v2}
\end{equation}
This equation is integrable for all $\alpha$.  However, the cases
$\alpha=0$ and $\alpha=1$ are sufficient for our purposes here.  We
note that this equation depends only on the fraction of the energy
deposited in each of the phases, and not on the form of the momentum
coupling.

By including the momentum coupling, the full shock structure as a
function of $x$ can be derived for each solution of
equation~\refeq{e:v1v2}, using
\begin{equation}
\rdif{x}{v_2} = -{\Phi_2 \over {\cal S}}.\label{e:xv2}
\end{equation}
Since ${\cal S}$ is a function only of local conditions in the flow,
it is clear that $v_2$ must decrease monotonically through the shock
structure (unless ${\cal S} = 0$ somewhere within the structure).

Dividing equation~\refeq{e:v1v2} by equation~\refeq{e:xv2}, we find
that a critical point exists, as expected, when the flow is at the
hot-phase sound speed, i.e., $v_1^2=\gamma p/\rho_1$, or
\begin{equation}
v_1 = {\gamma\over\gamma+1}{\Pi-\Phi_2 v_2\over \Phi_1}.\label{e:sonic}
\end{equation}
It may be possible for a continuous solution to pass through this
critical point if ${\cal S} = 0$, or if
\begin{equation}
v_2 = \left(1+{1\over\alpha(\gamma-1)}\right)v_1\label{e:crit2}
\end{equation}
(for $\alpha\ne0$).

\subsection{Shock solutions}

Shock solutions for the basic equations exist where the flow velocity
into the discontinuity exceeds the sound speed in the well-coupled
limit, \ie\
\begin{equation}
v|i > c^\star \equiv \left(\gamma
	p|i\over\rho|{1,i}+\rho|{2,i}\right)^{1/2}, \label{e:cstar}
\end{equation}
where $\rho|{1,i}$ and $\rho|{2,i}$ are the upstream densities, $p|i$
is the upstream pressure, and $v|i = v|{1,i} = v|{2,i}$ is the
upstream velocity.  

For $c^\star < v|i < c|{1,i} \equiv (\gamma p|i/\rho|{1,i})^{1/2}$,
the shock structures will be continuous in both fluids (C-type).  For
higher velocities the shocks will be led by viscous subshocks in the
hot gas (J-type) \cite[following the nomenclature defined for shock
waves with magnetic precursors by]{draine80}: as the hot-phase sound
speed is the fastest characteristic speed in the undisturbed gas, the
subshock must occur at the start of the resolved shock structure.  For
J-type shocks, the velocity of the hot phase just after the subshock
is
\begin{equation}
v|s = {\gamma-1\over\gamma+1}v|i+{2\gamma\over\gamma+1}{p|i\over\Phi_1},
\end{equation}
and the pressure is $p|s = p|i+\Phi_1(v|i-v|s)$ -- the standard
Rankine-Hugoniot conditions.

For $\alpha = 1$, energy conservation gives that the final velocity,
$v|f$, of a shock with upstream velocity $v|i$ and pressure $p|i$ is
given by
\begin{equation}
v|f = {\gamma-1\over\gamma+1}v|i+{2\gamma\over\gamma+1}{p|i\over\Phi_1+\Phi_2},
\end{equation}
whether or not it contains a viscous subshock.

For $\alpha = 0$, the situation is more complex.  For C-type shocks,
$v|f$ is given by
\begin{equation}
p|i\left(v|i\over v|f\right)^\gamma +(\Phi_1+\Phi_2)v|f = 
	p|i + (\Phi_1+\Phi_2)v|i.
\end{equation}
For J-type shocks with $v|i>c|{1,i}$, however, the final velocity,
$v|f$, is given in terms of $v|s$, $p|s$ behind the subshock by
\begin{equation}
p|s\left(v|s\over v|f\right)^\gamma +(\Phi_1+\Phi_2)v|f = 
	p|s + \Phi_1v|s+\Phi_2v|i.
\end{equation}
The expressions for $v|f$ are different for C- and J-type shocks for
all $\alpha \ne 1$.

Note that no shock solution can reach a sonic point where
equations~\refeq{e:sonic} and~\refeq{e:crit2} are satisfied.  We can
see this because in order to satisfy these conditions,
\begin{eqnarray}
v_2 = v_2^\star &\equiv& \left[p|i + (\Phi_1+\Phi_2)v|i\right]/\nonumber\\
&&	\left[\Phi_1\left(1+\gamma\over\gamma\right)
	\left(\alpha(\gamma-1)\over 1+\alpha(\gamma-1)\right) +
	\Phi_2\right].
\end{eqnarray}
If $\gamma>1$ and $0<\alpha\le1$, $v_2^\star > v|i$, but $v_2$ must
decrease monotonically through the shock structure [see the discussion
after equation~\refeq{e:xv2} above] so no such critical point can
occur in practice.

\subsection{Drag force}

\begin{figure*}
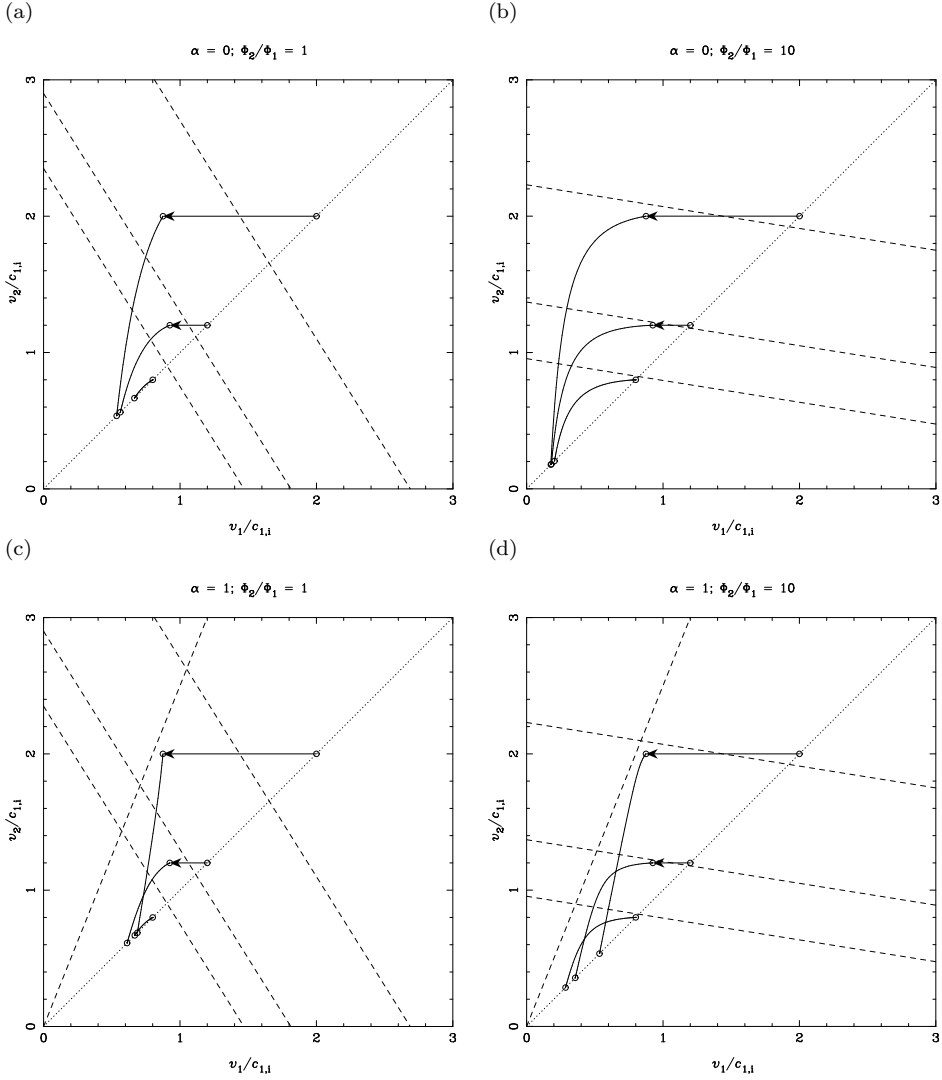

\begin{centering}
\begin{tabular}{ll}
(a) & (b) \\
\epsfysize=6cm\rotatebox{270}{\epsffile{f1a.ps}} &
\epsfysize=6cm\rotatebox{270}{\epsffile{f1b.ps}} \\
(c) & (d) \\
\epsfysize=6cm\rotatebox{270}{\epsffile{f1c.ps}} &
\epsfysize=6cm\rotatebox{270}{\epsffile{f1d.ps}} \\
\end{tabular}
\end{centering}
\caption{Variation of $v_1$ and $v_2$ through resolved shocks for
$\alpha=0$, $1$ and density ratios $\Phi_2/\Phi_1=1$, $10$.
Structures for shock Mach numbers $v|i/c|{1,i}= 2, 1.2, 0.8$ are shown
on each plot.  Circles show the initial and final states and the solid
curve the resolved structure, while the arrow shows the viscous
subshock in the structure if one is present.  The dotted curve shows
the equilibrium condition $v_1 = v_2$, while the dashed lines show the
critical conditions $v_1 = c_1$ (for each value of $v|i/c|{1,i}$) and,
for $\alpha = 1$, equation~\protect\refeq{e:crit2}.}
\label{f:shodo}
\end{figure*}

The spatial structure of the shocks which we calculate is determined
by the drag force, ${\cal S}$.  The simplest assumption is a relation
which takes the form of Stokes' law drag \cite{batch67}, for which
\begin{equation}
{\cal S} = \lambda\rho_1\rho_2(v_2-v_1).\label{e:ssimple}
\end{equation}
If the cool phase consisted of isolated, rigid, spherical clouds, then
$\lambda = 6\pi R|c \nu/M|c$, where $R|c$ and $M|c$ are the radius and
mass of the clouds and $\nu$ is the kinematic viscosity of the
intercloud medium.  Interpreted in this direct sense, this law is only
valid where the Reynolds numbers $2 R|c \vert v_2-v_1\vert/\nu \la 1$,
which will not apply in astrophysical situations: at larger Reynolds
numbers, the effective drag coefficient increases.  While it remains
of interest to consider this Stokes' drag force for an arbitrary drag
coefficient, the second order term in the expansion for the drag force
will cancel the dependence on flow viscosity but introduce a higher
order dependence on the slip velocity.

Assuming that the second-order term dominates leads to a drag law
proportional to the square of the flow velocity, as used by Shu
\etal~\shortcite{shuea72}, a dependence which also corresponds to the
`ballistic' case, where the drag force is calculated by assuming
independent collisions of hot-phase particles with the dense clouds.
It agrees well with empirical data for rigid projectiles.  In the
limit which we are considering, of small filling factor of clouds, the
drag force may be written
\begin{equation}
{\cal S} = C \left(9\pi\gamma^2\over128\right)^{1/3}
\rho_2(v_2-v_1){\rho_1^{1/3}\over M|c^{1/3}} \left(c_2\over c_1\right)^{4/3}
\vert v_2-v_1\vert,
\end{equation}
where the typical mass of a cold clump is $M|c$ and the typical
internal isothermal sound speed is $c_2$.  As above, $c_1$ is the
adiabatic sound speed in the diffuse medium.  

An alternative form for the drag force is based on the work of
Hartquist \etal~(1986, hereafter HDPS), who took into account the
effects of deformation of the clouds.  These authors considered the
dynamics of a flow around a clump in both the subsonic and supersonic
limits, assuming that the interaction zones were small enough that
each clump interacted with a smooth upstream flow.  They determined
forms for the mass-loss rate of the clumps, which have been used in
subsequent papers to predict a smoothed mass-input rate for the global
flow, and agree well with the numerical results of Klein
\etal~\shortcite{kmc} for strong shocks propagating over isolated
clouds.  It is a simple matter, however, to reinterpret their results
as a momentum transfer: the value is just the mass loss rate between
the phases derived by HDPS multiplied by the slip velocity.  If the
slip speed is slower than the sound speed in the hot phase, we find
that the coupling force is given by
\begin{eqnarray}
{\cal S} \simeq \rho_2(v_2-v_1) 
\left(\rho_1\vert v_2-v_1\vert c_2^2\over M|c\right)^{1/3}
{\rm min}\left((\vert v_2-v_1\vert/c_1)^{4/3},1\right).
\label{e:shdps}
\end{eqnarray}
We see that in both the Shu \etal{} and HDPS formulae, the momentum
transfer rate is proportional to the mass density in the cool phase
divided by the clump mass to the power $1/3$: if the cold gas is
placed in larger clumps, the ensemble is more permeable to the hot
phase flow.

It is of interest to study the results of these forms of slip term in
the test-particle limit (i.e.\ to treat the case in which phase 2
corresponds to a single cloud, and the hot phase velocity is
effectively constant).  From equations~\refeq{e:2mass}
and~\refeq{e:2mom}, we find that the velocity of the cloud relative to
the hot phase gas is
\begin{equation}
\rdif{v_2}{t} = - {{\cal S}v_2\over\Phi_2} = -\Lambda v_2^\beta.
\end{equation}
For Stokes' drag, $\beta = 1$ and $\Lambda = \lambda\rho_1$, while for
the (subsonic) HDPS law, $\beta = 8/3$ and $\Lambda$ is a constant
independent of $v_2$.  Integrating, we find that for $\beta = 1$,
\begin{eqnarray}
v_2 & = & v|i \exp(-\Lambda t)\\
x   & = & {v|i\over\Lambda}\left[1-\exp(-\Lambda t)\right],
\end{eqnarray}
for $\beta = 2$, 
\begin{eqnarray}
v_2 & = & {v|i\over 1+\Lambda v|i t}\\
x   & = & {1\over\Lambda}\log(1+\Lambda v|i t),
\end{eqnarray}
and otherwise
\begin{eqnarray}
v_2 & = & v|i\left[1+(\beta-1)\Lambda v|i^{\beta-1} t\right]
	^{-1/(\beta-1)}\\
x   & = & {v|i^{2-\beta}\over(\beta-2)\Lambda}
	\left\{\left[1+(\beta-1)\Lambda v|i^{\beta-1} 
	t\right]^{(\beta-2)/(\beta-1)}-1\right\}.
\end{eqnarray}
In the Stokes' drag case, the dense cloud tends to a finite
displacement, $v|i/\lambda\rho_1$, from its initial position in the
diffuse flow.  However, for the subsonic HDPS law (and in general for
all $\beta \ge 2$) there is no such limit.  

This observation relates to d'Alembert's paradox -- the prediction of
zero drag on a body in irrotational flow, contrasted to practical
experience in which turbulence in the wake of bluff bodies maintains
drag.  For slender bodies in flows of high Reynolds number, very low
drag is observed in reality \cite{batch67}, and so long slip distances
could indeed occur for in some cases.  We include results for both
types of behaviour in what follows.

\section{Shock structures}

\begin{figure*}
\begin{centering}
\begin{tabular}{ll}
(a) & (b) \\
\epsfysize=8cm\rotatebox{270}{\epsffile{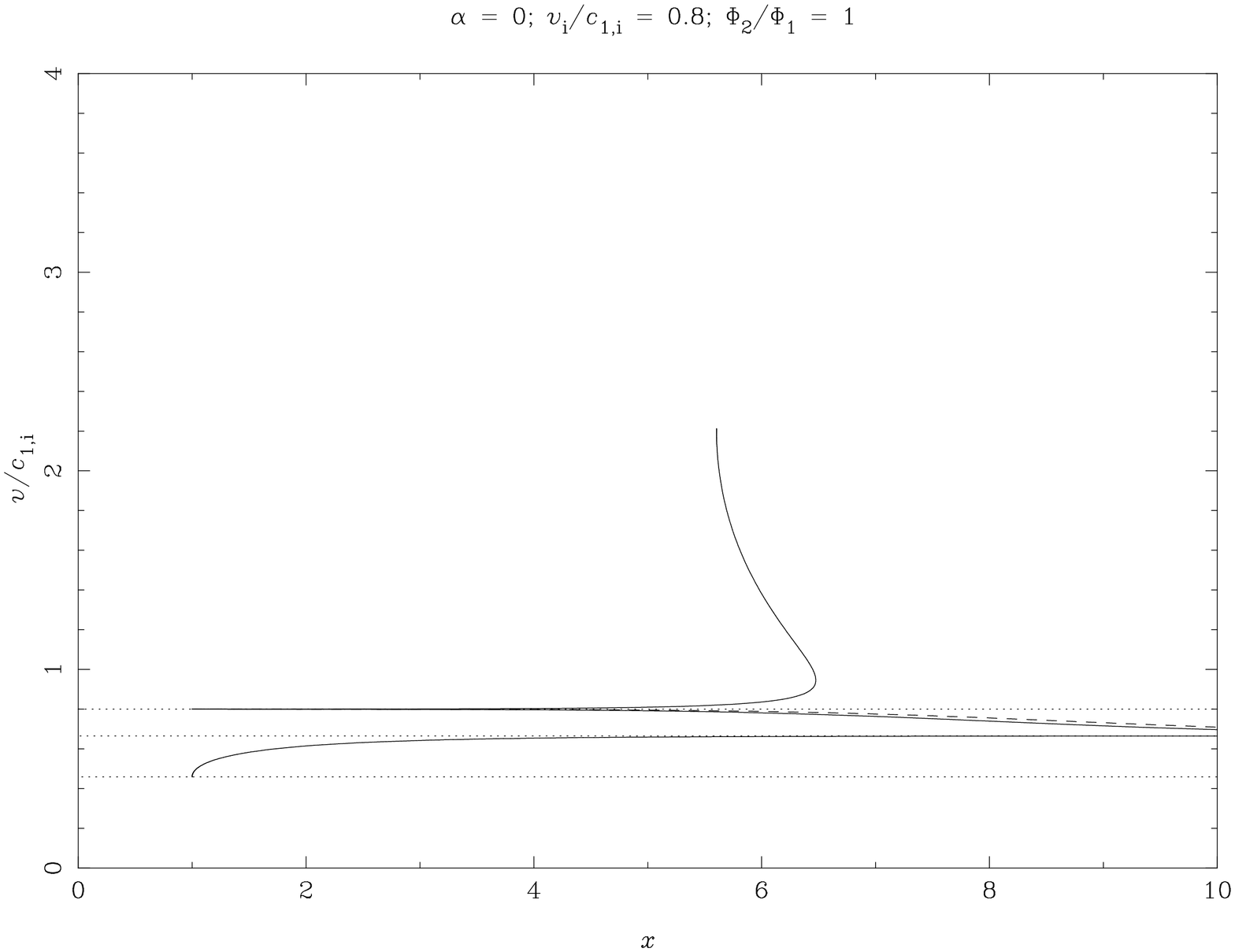}} &
\epsfysize=8cm\rotatebox{270}{\epsffile{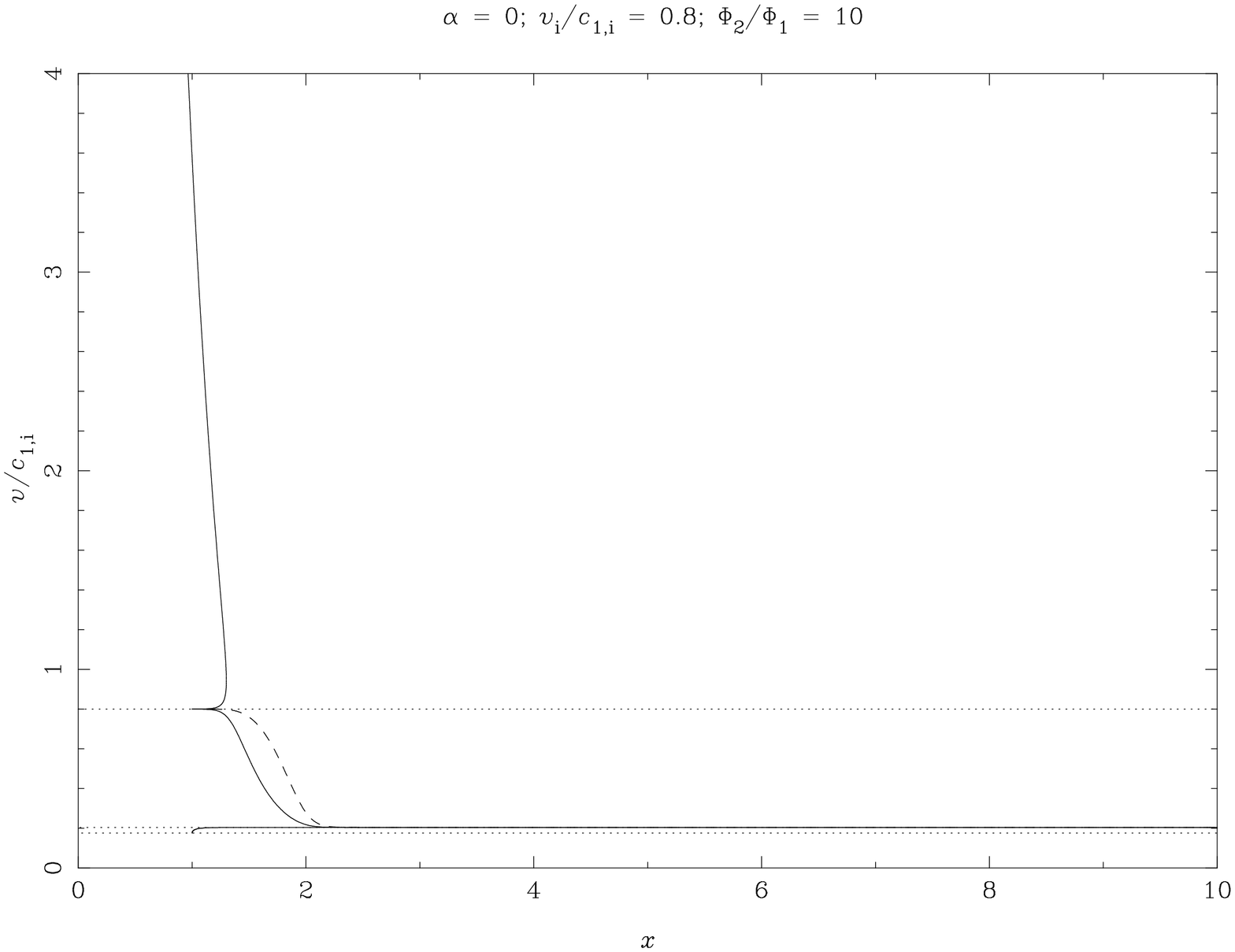}} \\
(c) & (d) \\
\epsfysize=8cm\rotatebox{270}{\epsffile{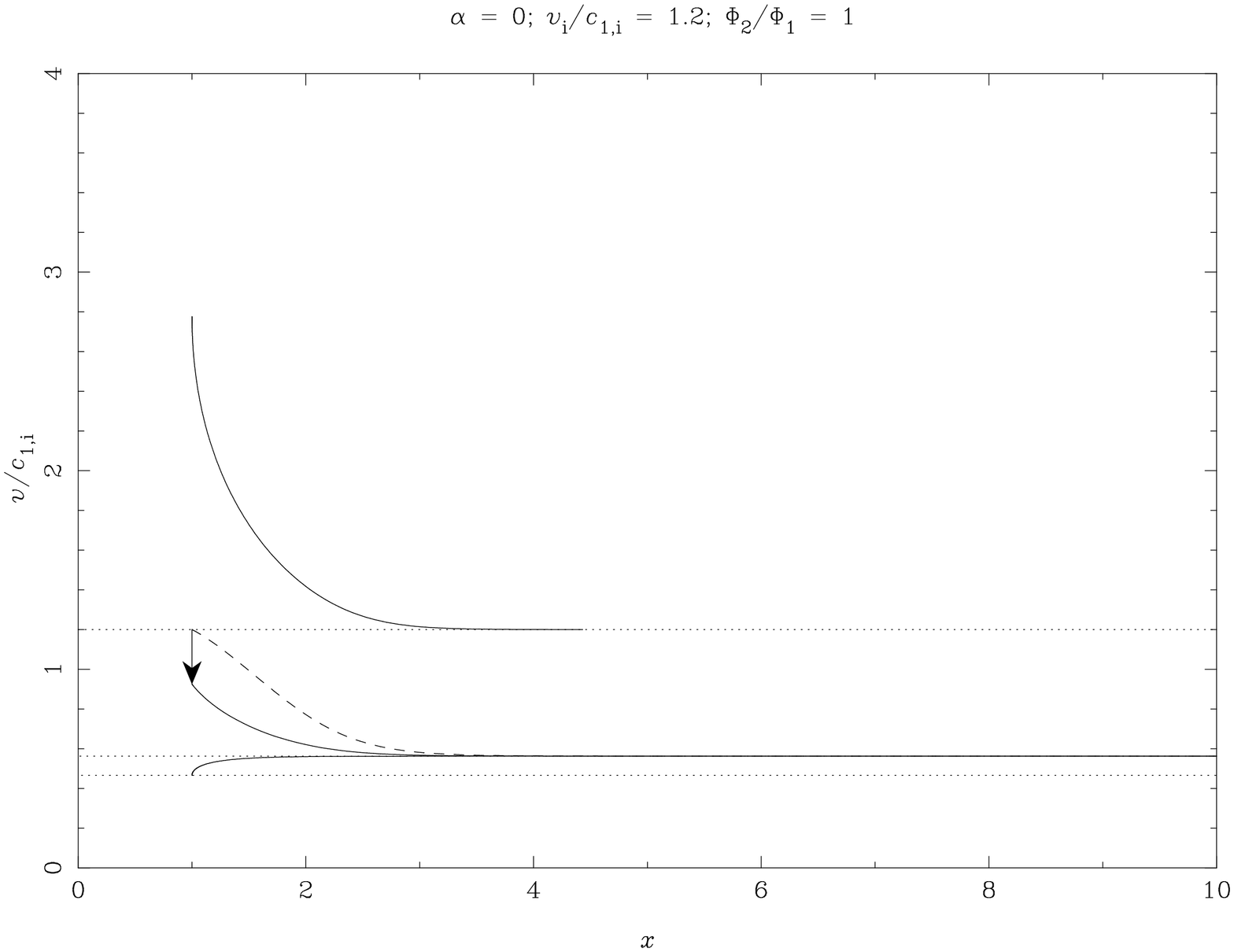}} &
\epsfysize=8cm\rotatebox{270}{\epsffile{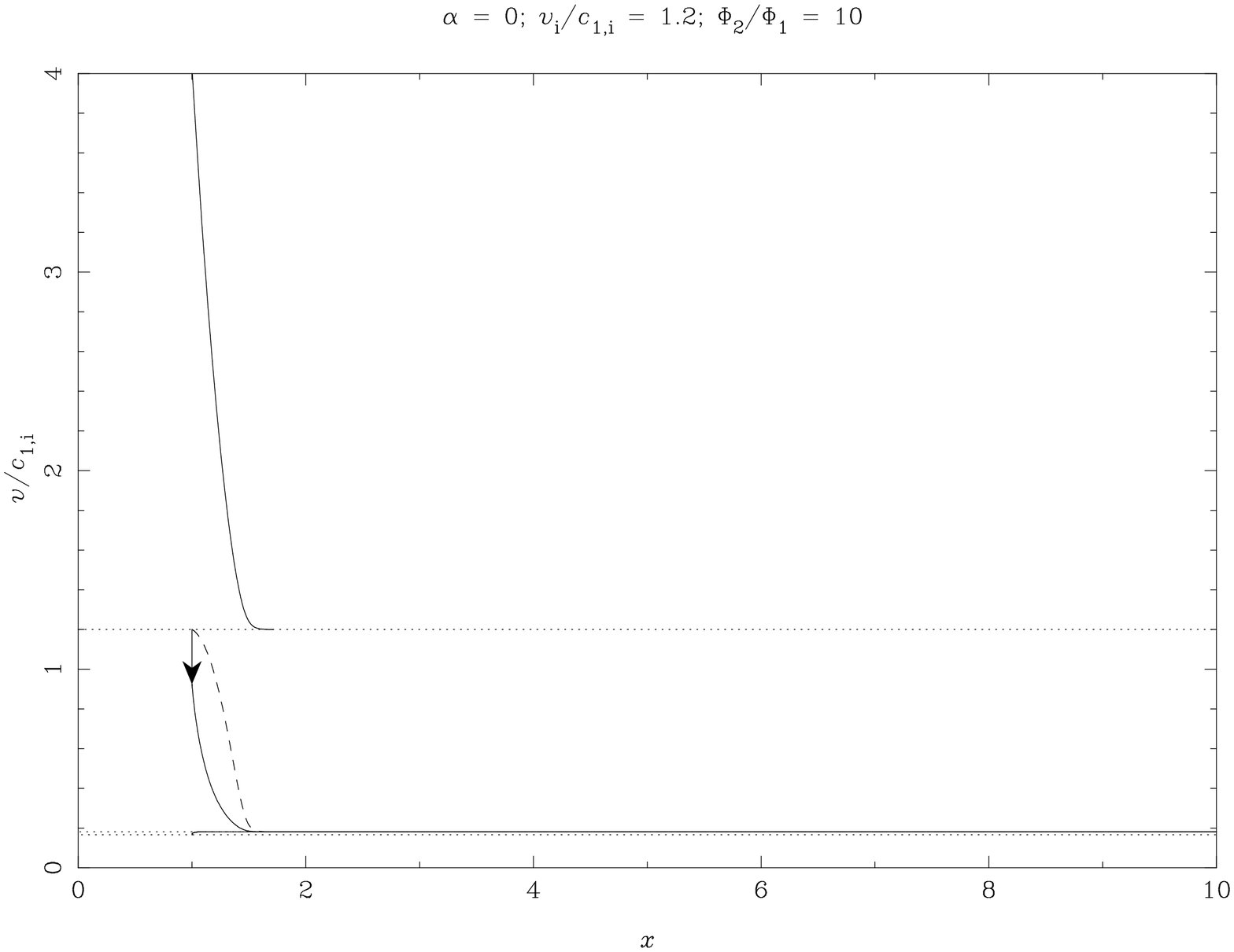}} \\
(e) & (f) \\
\epsfysize=8cm\rotatebox{270}{\epsffile{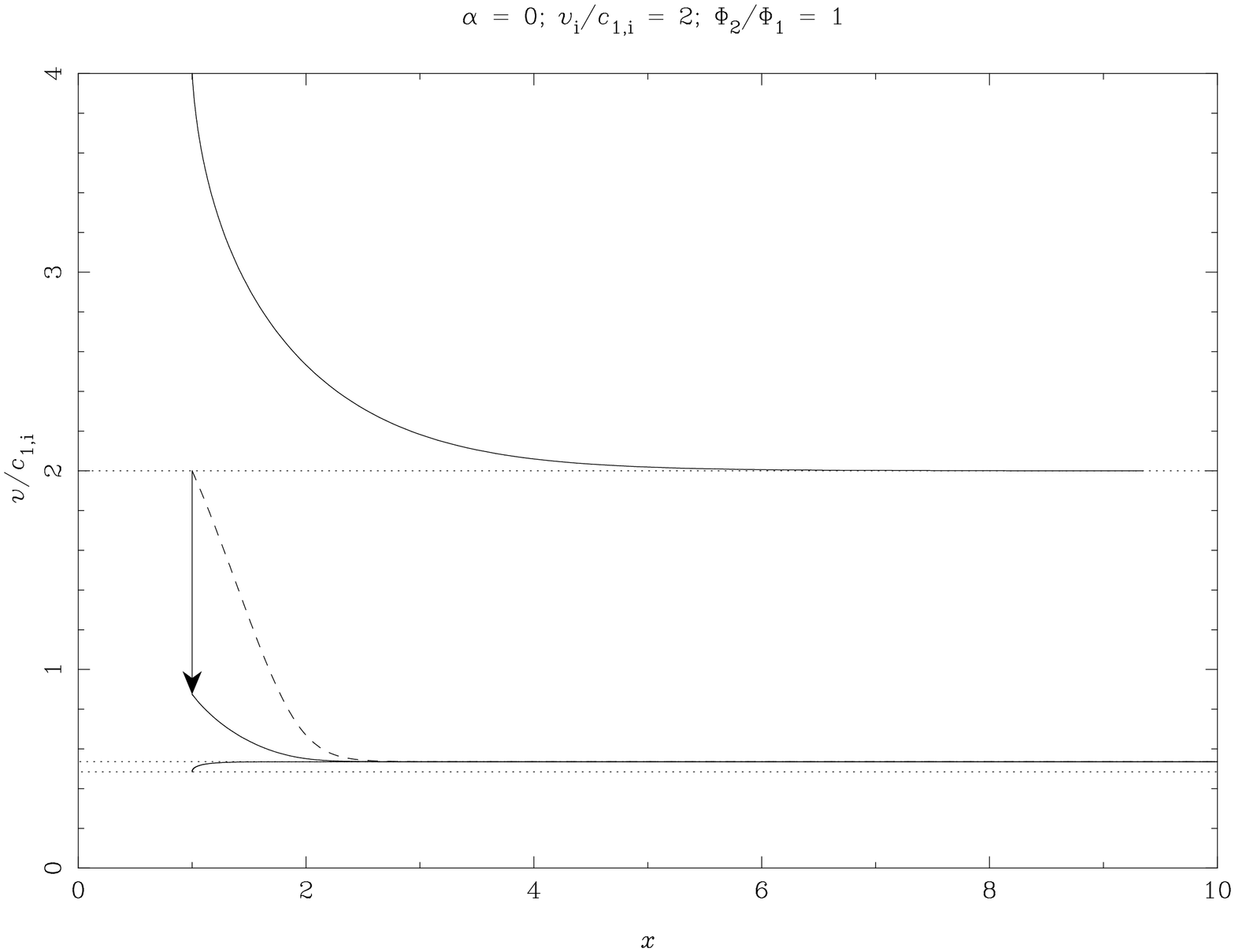}} &
\epsfysize=8cm\rotatebox{270}{\epsffile{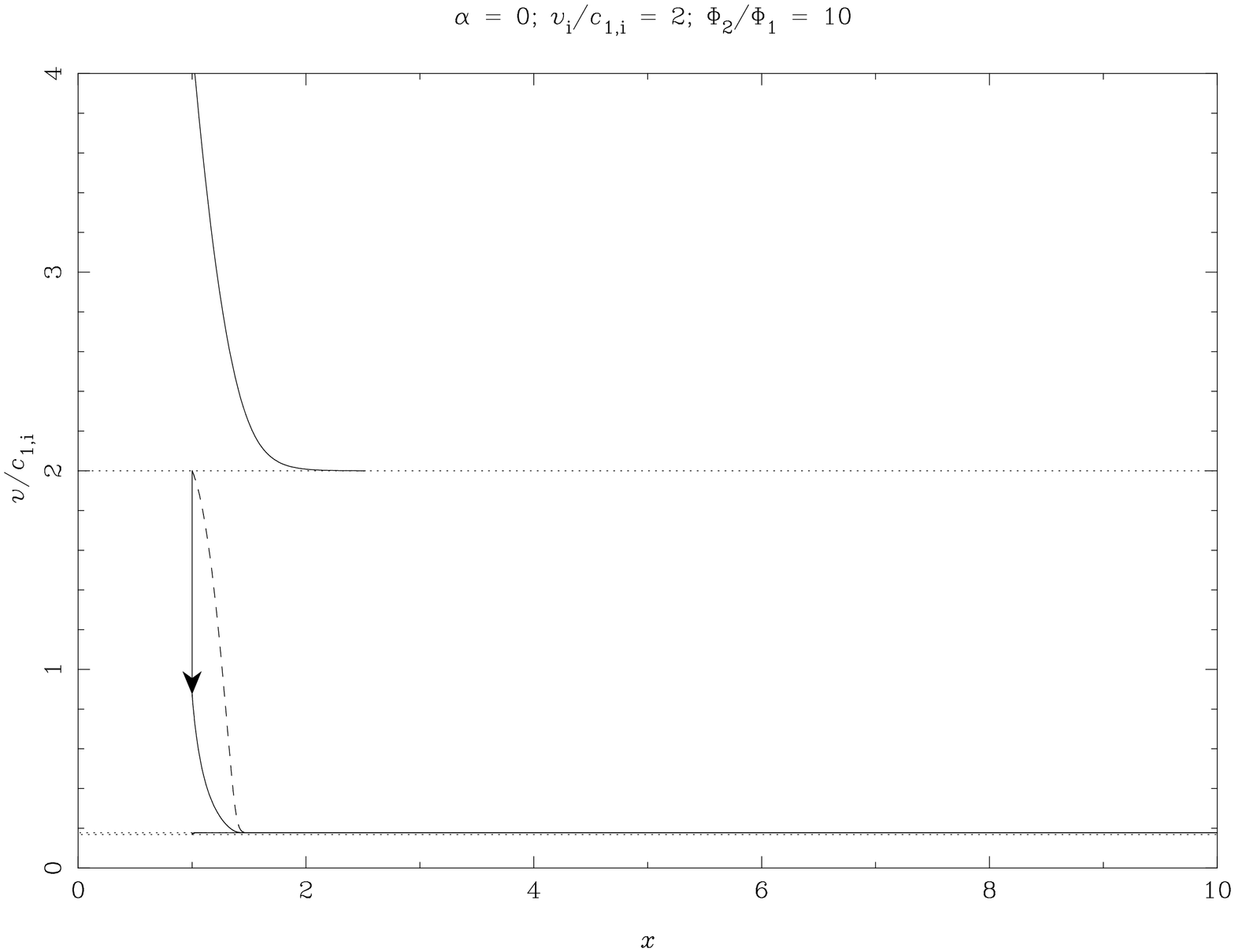}} \\
\end{tabular}
\end{centering}
\caption{Shock structures for $\alpha=0$ and various shock Mach
numbers $v|i/c|{1,i}$ and density ratios $\Phi_2/\Phi_1$ and with a
slip force given by equation~\protect\refeq{e:ssimple}, for $c|{1,i} =
\rho|{1,i} = 1$. The dotted curves show the initial and final
velocities of the shocks, as well as $v_1$ for a limiting case with
$v_2 = 0$ which relaxes to the same final state.  The solid curve
between the upper dashed lines shows the variation of $v_1$ through
the shock (including the initial subshock for $v|{1,i}>c|{1,i}$),
while the dashed curve shows $v_2$.  Additional solid curves show
solutions relaxing to the pre-shock and post-shock equilibrium
states.}
\label{f:sstruc0}
\end{figure*}

\begin{figure*}
\begin{centering}
\begin{tabular}{ll}
(a) & (b) \\
\epsfysize=8cm\rotatebox{270}{\epsffile{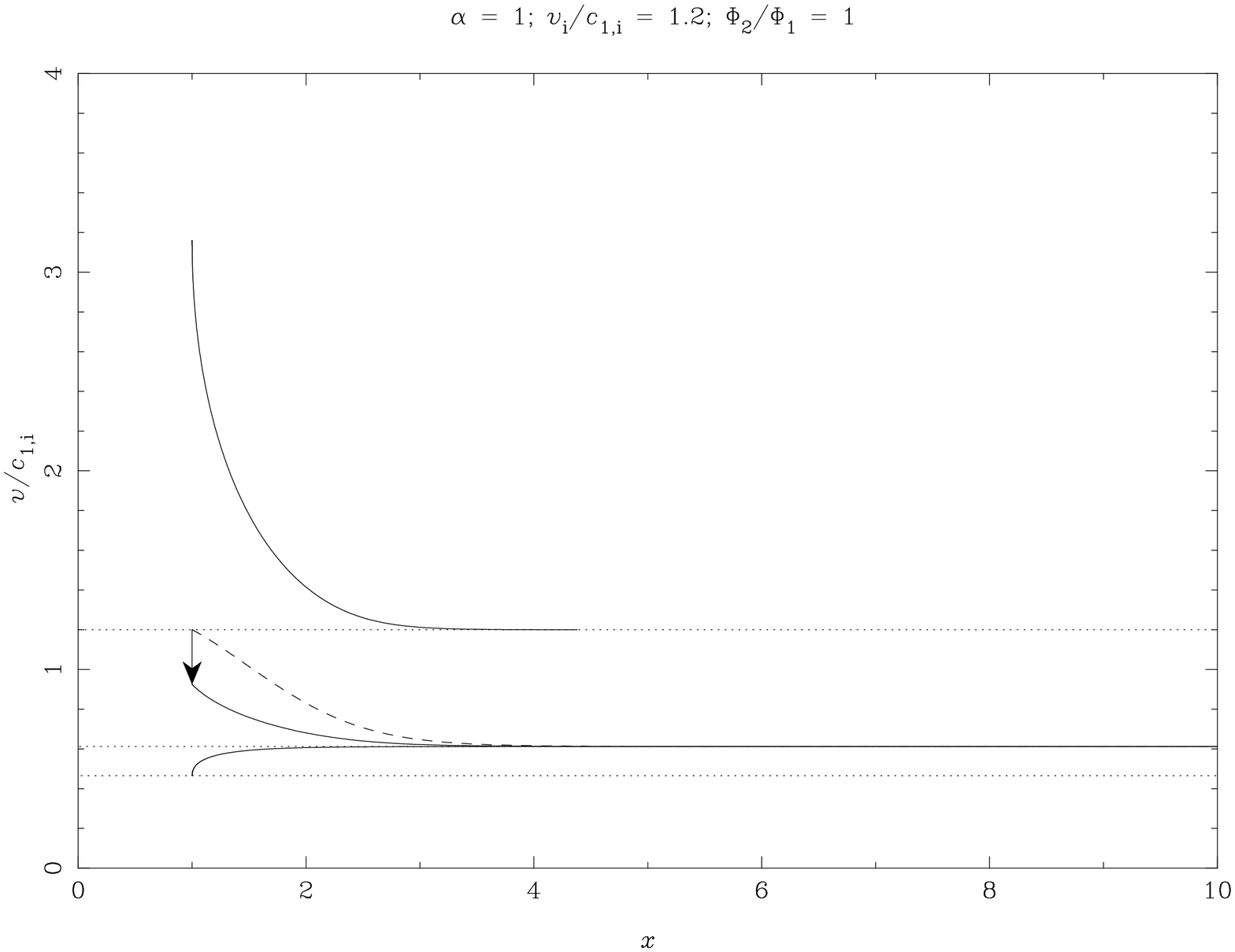}} &
\epsfysize=8cm\rotatebox{270}{\epsffile{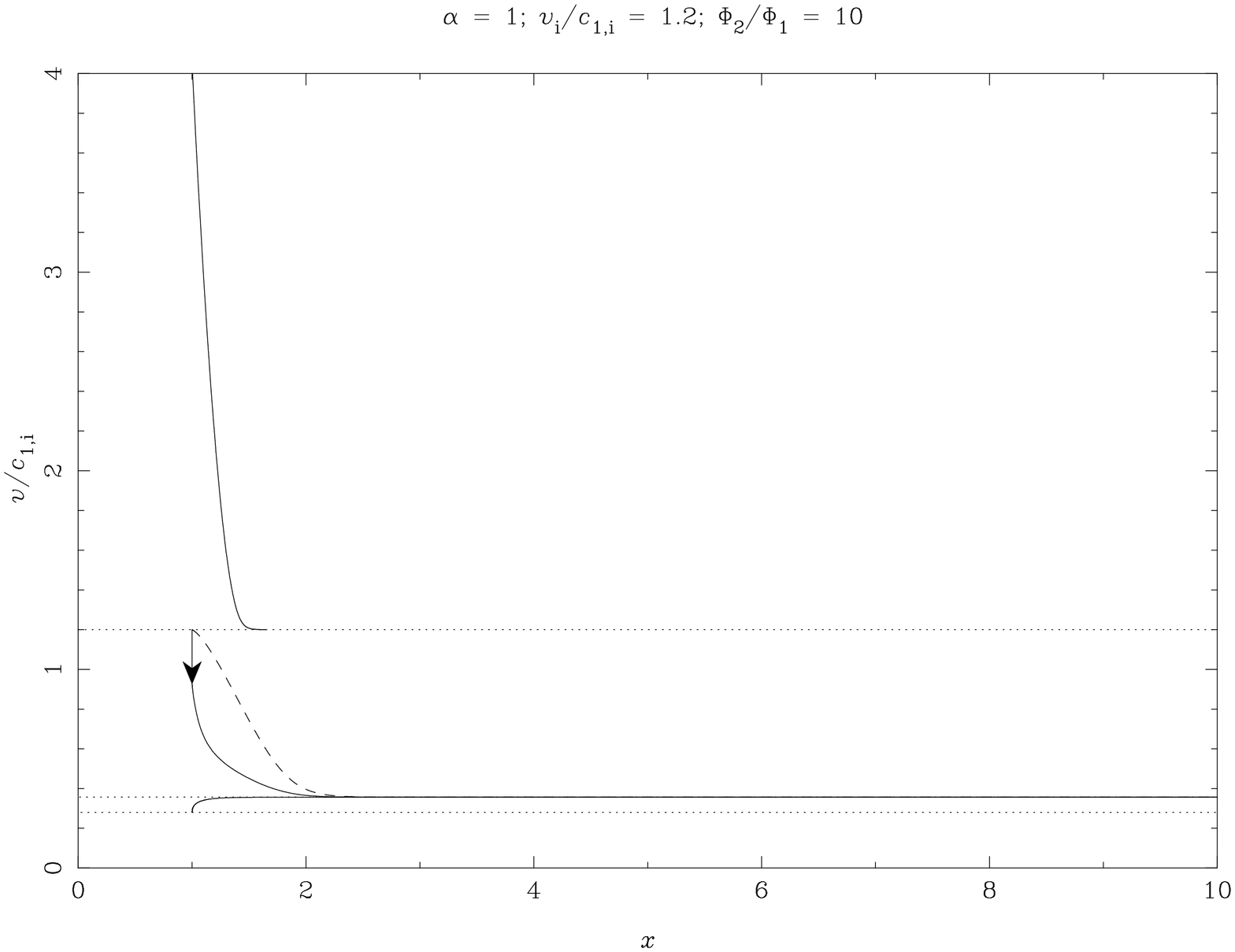}} \\
(c) & (d) \\
\epsfysize=8cm\rotatebox{270}{\epsffile{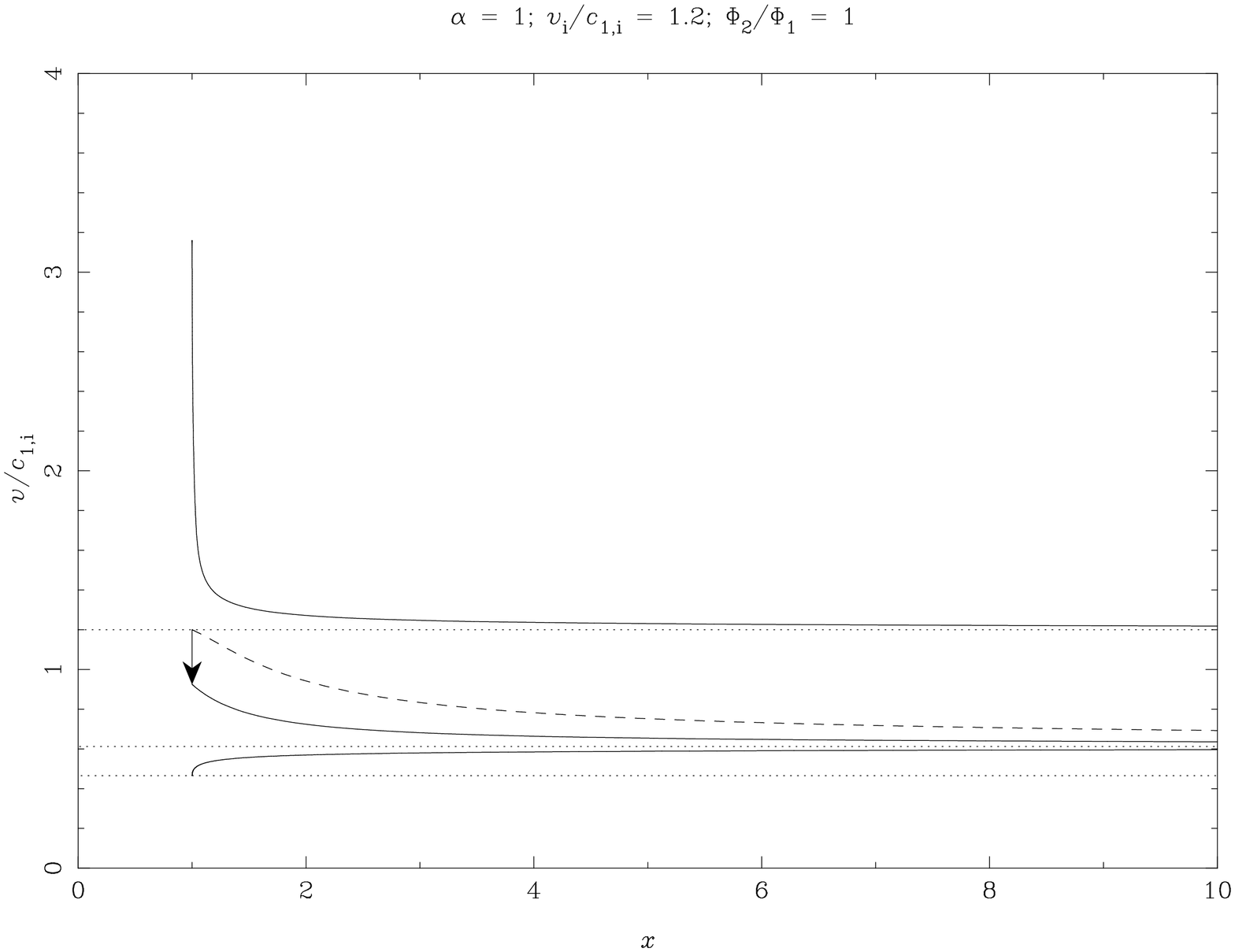}} &
\epsfysize=8cm\rotatebox{270}{\epsffile{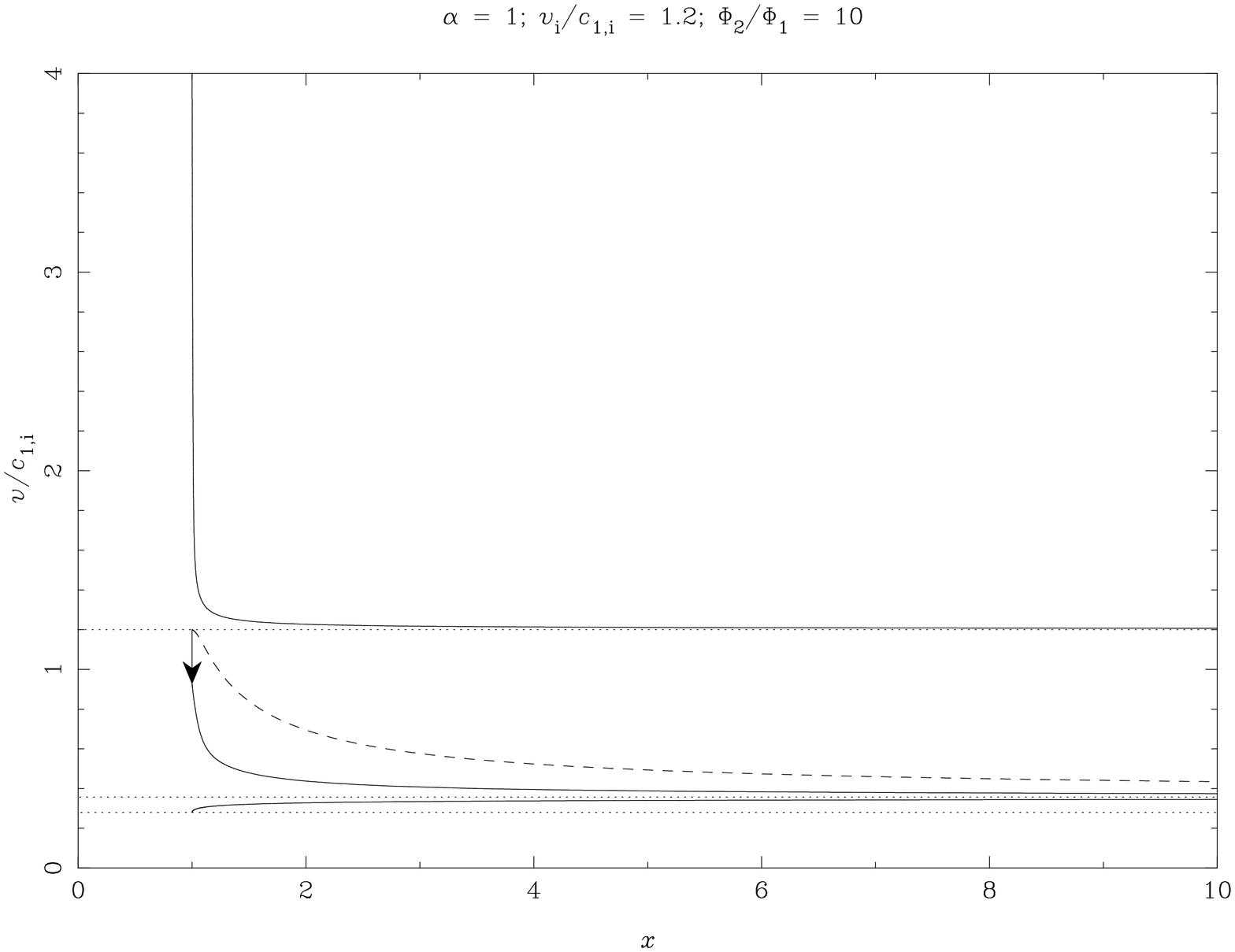}} \\
\end{tabular}
\end{centering}
\caption{Shock structures for $v|i/c|{1,i} = 1.2$ as in
Figure~\protect\ref{f:sstruc0}, except that $\alpha=1$, and for (c),
(d) the HDPS slip law is used.}
\label{f:sstruc1}
\end{figure*}

In Figure~\ref{f:shodo}, we plot the structure of the shocks in
drag-coupled two phase media in the $(v_1,v_2)$ plane, for $\alpha =
0$ and $\alpha = 1$ respectively.  These plots are independent of the
form of the drag force, ${\cal S}$ [\cf\ equation~\refeq{e:v1v2}].
The two columns of these Figures are for a density ratio of
$\rho|{1,i}/\rho|{2,i}$ of $1:1$ and $1:10$, respectively, with curves
for several shock Mach numbers (with respect to the adiabatic sound
speed in the smooth phase, $c|{1,i}$) shown on each plot.  For
comparison, for a $1:1$ density ratio, the equilibrium critical speed
is $c|{1,i}/2^{1/2}$, while for a $1:10$ ratio is is
$c|{1,i}/11^{1/2}$.  For $\alpha=0$, the exhaust velocity of the
shocks decreases for the strongest shocks, while for $\alpha=1$, the
exhaust velocity is increased as a result of the energy deposition in
the hot phase.

For the weakest shocks, we see that the velocities in the two phases
remain very similar, and have continuous (C-type) profiles.  For case
$\Phi_2/\Phi_1 = 10$, substantial compressions of the medium can be
achieved in these C-type shocks.  As the shocks strengthen, the smooth
phase begins to decrease in velocity significantly before the clumped
phase.  

Once $v|i > c|{1,i}$, a viscous subshock is required in the solutions
(shown as an arrow), which are hence J-type.  It is apparent that, for
$\alpha = 0$, increasing clumped-phase density results in a stronger
shock (for fixed $v|i/c|{1,i}$), and more post-shock velocity
structure in the smooth phase gas.  These effects are less marked for
$\alpha = 1$.  For shocks with $\alpha = 1$ stronger than those shown
here, the smooth-phase gas accelerates after passing through the
subshock.

In Figures~\ref{f:sstruc0} and~\ref{f:sstruc1}, we show the internal
structure of the shocks as a function of distance, for the Stokes'
drag law, equation~\refeq{e:ssimple}, and for the HDPS slip law,
equation~\refeq{e:shdps}.  The shock structures get narrower as the
shock velocity increases, while going from $\alpha=0$ to $\alpha=1$
broadens them somewhat.  Increased mass flux in the cold phase,
$\Phi_2$, appears to narrow the shocks.  This is a result of the
reduction of the critical speed for the well-coupled limit, which
means that for fixed velocity through the upstream gas, the shocks
with higher $\Phi_2$ are stronger.  Note in particular the extremely
small slip velocity between the phases throughout the C-type shock
shown in Figure~\ref{f:sstruc0}(a).

The dashed lines in the Figures show the variation of $v_2$ through
the shock: it is apparent that the velocity difference between the
phases increases for stronger shocks.  Additional solid lines show
supersonic and subsonic equilibria.  For shock speeds smaller than the
hot-phase sound speed, the solutions of the o.d.e. diverge from the
higher-velocity equilibrium solution.  As we have previously discussed
\cite{wd96}, the higher-velocity equilibrium solution is not
physically unstable, but this solution topology will occur if fully
resolved shock structures are present.

Comparing the results for Stokes' and HDPS drag, we have set the
constants of proportionality so the initial post-shock relaxation
occurs at a similar rate.  It is apparent that the HDPS form leads to
an extended post-shock relaxation region, as the small slip velocity
in this region leads to decreased effective friction.

\subsection{Vorticity}

The question of whether the passage of a shock through a clumpy medium
smooths out the clumpy structure depends largely on transport
processes such as viscosity and thermal conduction (although in
astrophysical plasmas, radiative processes can also play an important
role).  The vorticity produced by the passage of the shock is crucial
in enhancing transport processes, as a result of the formation of
viscous boundary layers.

If there is a viscous subshock, then the flow is not homoentropic and
so Kelvin's circulation theorem (e.g.\ Batchelor 1967, p273--277) does
not hold.  Shocks in the flow will generate vorticity when their
strength changes across their surfaces.  Considering the small scale
structures of the flow as the subshock propagates through it, it is
apparent that vorticity will be generated in the hot phase where the
incident shock refracts around obstacles, and by reflected shocks
ahead of them.  The forward shock driven into each clump will also
lead to creation of vorticity within the clump.  This wide
distribution of vorticity within the flow will have a significant
disruptive effect~\cite[\cf{}]{kmc,jjn}.

However, if there is no viscous subshock and $\alpha = 0$, the
hot-phase flow is homoentropic and (as a result of Kelvin's
circulation theorem), the vorticity produced will be entirely confined
to the cold phase and the contact between the phases.  Even
for $\alpha \ne 0$, the generation of vorticity should be confined to
the regions of the hot phase flow close to the boundary layers in
which the frictional heating is deposited.

As the slip velocity gets smaller, the shear velocities over the
surface of individual clumps decrease, and so does the effectiveness
with which cool material is entrained from their surfaces.  Instead,
the cool clumps are rather gently accelerated as they cross the broad
shock structure.

We can develop various approximate measures of the destructiveness of
the boundary layers induced by passage through a shock.  For instance,
the amount by which two initially coincident particles, one in each
phase, are dragged apart by the front,
\begin{equation}
\Delta x = v|f\int_{-\infty}^\infty {1\over v_1}-{1\over v_2} dx,
\end{equation}
where $v|f$ is the final equilibrium velocity behind the shock, and
$v_1$ and $v_2$ are the velocities of the different phases as they
pass through it.  The value of $\Delta x$ may be compared to other
characteristic lengthscales in the problem.  If it is smaller than the
size of an individual cold clump, then the passage of the shock will
have very little effect on the structure of the medium, while if it is
larger than the inter-clump separation, the cool phase will likely be
dispersed into a fine aerosol.

Using equation~\refeq{e:2mom} and the Stokes' drag force,
equation~\refeq{e:ssimple}, we find that
\begin{equation}
\Delta x = {v|f\over\lambda\rho_{1,i}}\left(1-{v|f\over v|i}\right).
\end{equation}
It is clear that as the shocks weaken or the coupling improves,
$\Delta x$ becomes smaller.  By contrast, the slip law of HDPS leads
to increasingly poor coupling between the phases at small slip
velocities compared to Stokes' drag and the slip distance would be
expected to increase for small velocity jumps \cite[\cf{}]{yw82}.

An alternative measure of destructiveness is the maximum slip velocity
between the phases.  This can be compared, for instance, to the
peculiar velocities of the clumps in the upstream multiphase gas: if
the maximum slip velocity is smaller than this value, it is difficult
to see why the multiphase structure should be substantially effected
by the passage of the shock.  The interstellar medium has significant
structure on large scales, while for a purely hydrodynamic flow in
radiative equilibrium, it would be expected that different phases
would be dispersed as an aerosol.  However, scale lengths are
introduced by effects of such as self-gravity (the Jeans length),
thermal conduction [the Field~\shortcite{field65} length] or
photoionization (the recombination length), resulting in large-scale
clumping.  If these structures are to form, then they must have a
binding energy sufficient to outweigh the higher entropy of the
dispersed state.

It is clear, from Figure~\ref{f:shodo}, that the maximum slip velocity
decreases rapidly with shock strength.  For example, for $\alpha = 0$,
the maximum value of the slip velocity, $V=v_2-v_1$ is given by
\begin{equation}
V^\star = \left(\Phi_1+\Phi_2\over\gamma\Phi_2\right)
\left[\left(v_1^\star\over v|f\right)^{\gamma}+\gamma{v|f^\star\over
v_1} - (\gamma+1)\right]v_1,
\end{equation}
for a shock with exhaust velocity velocity $v|f$, and velocity
$v_1^\star$ at the maximum of $V^\star$.  As $v_1^\star \to v|f$,
$V^\star = O[(v|f-v_1^\star)^2]$, so for weak shocks the maximum slip
velocity decreases rapidly.

If the square of the maximum slip velocity is less than the effective
binding energy per unit mass of these structures, then the shock
should be able to change the velocity of the flow as a whole without
greatly effecting its multiphase structure.  While some details of the
shock structures vary substantially between different forms of the
momentum coupling law, it is in all cases that the maximum slip
velocity decreases rapidly as the overall shock velocities decreases
below the hot phase critical speed.

\section{Applications}

\begin{figure}
\epsfxsize=8cm\mbox{\epsffile{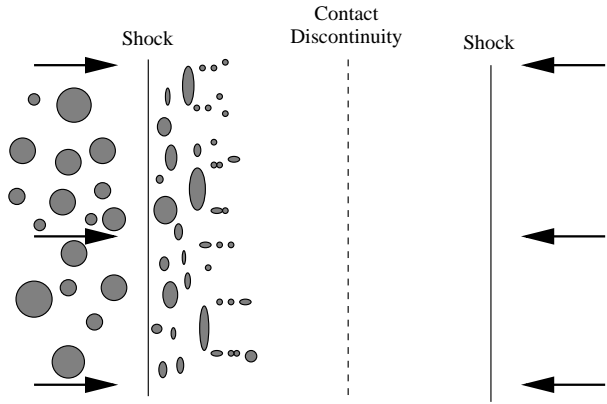}}
\caption{Schematic drawing of the structure of a multiphase shock.}
\label{f:riemann}
\end{figure}

The relatively benign environment of resolved shocks may have
important implications for the evolution of multiphase structures in a
wide range of astrophysical circumstances.  In the present section, we
outline some of those we expect, pending future more detailed
investigations.

In Figure~\ref{f:riemann}, we show a schematic picture of the
structure of shock-tube problem, in which two gas components collide.
The component moving in from the right is smooth, while the component
moving in from the left has a clumped structure.  If the flows from
either side are sufficiently rapid, the solution will take up the
usual form of reverse shock, contact discontinuity, forward shock.  In
Figure~\ref{f:riemann}, we illustrate the situation in which the
reverse shock into the multiphase medium is of J type: after they pass
through the shock, the clumps become elongated and are eventually
destroyed (although the deformation of the clumps will obviously be
far more complex than that shown in the Figure).  If the shock is of C
type, however, we have argued that the clumps will be far more
gradually decelerated and as a result could remain as fast-moving
coherent structures within the protective cocoon of the shocked gas
layers.

If a clump slips forwards through the contact discontinuity, its
environment will become somewhat more harsh.  This will occur if the
dense clumps slip through the hot material for a distance larger than
that between the shock and the contact discontinuity, or, in an
environment which is not plane-parallel, as a result of sideways
escape of the diffuse gas.  As the forward shock will often be curved,
the flow beyond the contact discontinuity will often have significant
shear flows.  Once a clump passes through the line of the forward
shock, however, it will shock directly against the supersonic
impinging material and will rapidly dissipate.  

This may help explain the survival of knots and bullets in regions
such as Orion \cite{tedds99} and Eta Carinae \cite{meab96}.  In
systems such as these, the spatial density of the bullets may be high
enough that they maintain a relatively shared bow-shock, and so reduce
the destructive vorticity of the post-shock flow.

It is generally thought that the structures of planetary nebulae are
formed when a hot, diffuse wind from a white dwarf remnant overtakes a
cool, dense stellar wind blown off during the stars previous evolution
along the asymptotic giant branch \cite{kwokea78,mell95}.  The cool
wind is likely to be highly structured, and clumps advected into the
hot bubble can have important effects on the global flow \cite{adh94},
as well as being observed in objects such as the Abell
30~\cite{borkea95}.  Before they start to interact with the hot
stellar wind, these clumps will have had to pass through an outer
shock without being destroyed.  In particular, the outer region of
shocked gas may be broadened substantially, compared to numerical
models which have often assumed the cool wind is uniform.

This analysis also provides insight into the structure of the flows
around nuclear starbursts (Strickland \& Stevens 2000, and references
therein).  Energy input from supernovae and stellar winds in these
starbursts drives plumes of hot gas perpendicular to the plane of the
host galaxy, over distances of many kiloparsecs.  In the plane of the
galaxy, the hot wind from the starburst interacts with the ISM.  To
date, simulations have in general assumed that the host galaxy ISM may
be treated as a smooth gas, at a temperature of $\sim 10^5\Kelv$.
This temperature is similar to the mass weighted mean random velocity
in the clumpy ISM, and so gives a fairly reasonable impression of the
coarse-grained structure of the ISM before the interaction with the
nuclear wind begins.  Strickland \& Stevens~\shortcite{stricks00} do
investigate the effects of mass-loading in some of their models, and
find relatively small effects on the flow structure. However, in the
galactic plane in their simulations the distributed gas has a
temperature of only $\sim 7\ee4\Kelv$, while we find here that the
hotter components of the gas will be very important for transmitting
pressure waves.

When the nuclear starburst starts to drive a flow outwards in the
simulations, the region around the starburst is first evacuated of the
ambient ISM gas.  After this, a nuclear wind develops which becomes
supersonic close to the edge of the starburst mass injection.  The
resulting nuclear wind shocks against the residual ISM with the usual
structure of a forward-shock, a contact discontinuity and a reverse
shock.  Some cool gas may be dragged away from the galactic plane by
shear flows along the contact discontinuity, but in general the
structures seen in the galaxy plane are very sharp and well defined.

In contrast, the broadening of the shocks will mean that clouds in the
ISM will be rather gradually driven away from the starburst by a
diffuse, broad outer shock.  The small force on the clouds will result
in a secular increase in orbit energy, rather than any dramatic
outflow.  As a result, in the plane of the galaxy the inner shock and
sonic transition around the starburst may take a considerable period
to develop.  The hot wind from the central nuclear starburst will be
surrounded by a weaker wind which has percolated through the ISM, and
which carries some shreds of advected cool material along with it.

\section{Conclusions}

In this paper, we have discussed the effects which the multiphase
nature of a medium can have on the structure of shocks passing through
the medium.  The shocks rapidly become broader, particularly when the
shock speed in the well-coupled limit is smaller than the sound speed
in the medium which fills most of the space.  We discuss how, in this
case, multiphase structure would be expected to survive the passage of
the shock.

Broadened multiphase shocks will be important in a wide variety of
astrophysical circumstances, as we have discussed above.  Equations
such as our equations~\refeq{e:1mass}--\refeq{e:2mom}, which model the
effects of small-scale structures, can be used in time-dependent
hydrodynamical modelling of large-scale structures.  For a small
computational expense, significant additional insights into the
structure of astrophysical flows are available, without the need to
resolve the structure of the flow to its finest details.

\section*{Acknowledgments}

RJRW wishes to thank the PPARC for their support through the award of
an Advanced Fellowship, and the Department of Physics and Astronomy,
University of Leeds for hospitality while this work was developed.

\label{lastpage}
\end{document}